# Structural and electronic properties of the metal-metal intramolecular junctions of single-walled carbon nanotubes

Wei Fa[1, 2], Jiangwei Chen[1], Hong Liu[1] and Jinming Dong[1]

1. *National Laboratory of Solid State Microstructures and Department of Physics, Nanjing University, Nanjing, 210093, People's Republic of China.*

2. *Department of Applied Physics, Nanjing University of Aeronautics and Astronautics, Nanjing, 210016, People's Republic of China*

**Abstract.** Several intramolecular junctions (IMJs) connecting two metallic (11, 8) and (9, 6) carbon nanotubes along their common axis have been realized by using a layer-divided technique to the nanotubes and introducing the topological defects. Atomic structure of each IMJ configuration is optimized with a combination of density-functional theory (DFT) and the universal force field (UFF) method, based upon which a four-orbital tight-binding calculation is made on its electronic properties. Different topological defect structures and their distributions on the IMJ interfaces have been found, showing decisive effects on the localized density of states, while the σ-π coupling effect is negligible near Fermi energy ($E_F$). Finally, a new IMJ model has been proposed, which probably reflects a real atomic structure of the M-M IMJ observed in the experiment [Science **291**, 97 (2001)].
.

## I. Introduction

Single-walled carbon nanotubes (SWNTs), which are graphite cylinders made of a hexagonal carbon-atom lattice, have drawn a great deal of interests due to their fundamental research importance and tremendous potential technical applications [1]. For example, they might play an important role in future molecular electronic devices, such as room-temperature single electron and field-effect transistors [2-3], and rectifiers [4-6]. A SWNT can be either a semiconductor or a metal, depending on its helicity and diameter [7-10]. Thus, connection of two segments of SWNTs with different diameters and helicities by introducing the pentagon-heptagon defects into the perfect hexagonal network will create the metal-semiconductor (M-S), semiconductor-semiconductor (S-S), or metal-metal (M-M) IMJ, which could be the building blocks for much smaller electronic devices than those with conventional semiconductor technology.

The electronic properties of the SWNT IMJs have been the subject of an increasing number of experimental and theoretical studies [11-12] since 1995. Chico *et al.* [13-14] indicated that a pentagon-heptagon pair with symmetry axis nonparallel to the tube axis could change the nanotube helicity by one unit from (n,m) to (n+1,m-1). J.-C.Charlier *et al.* [15] addressed general features of the IMJs in the zigzag configuration. Such kinds of studies suggested that these structures could function as nanoscale electronic devices made entirely of carbon atoms. Although a great progress in the IMJ research has been made, there are still a lot of experimental and theoretical uncertainties about the close relationships between their physical properties and corresponding geometrical structures, e.g., how to build the IMJs by the topological defects (pentagons, heptagons, 5-7 pairs, and their different combinations), and what are their effects on the local density of states (LDOSs) and other electronic properties of the IMJs?

Recently, Min Ouyang *et al*. [16] studied the (21, -2)/(22, -5) and (11, 8)/(9, 6) IMJs by using scanning tunneling microscopy (STE), which provided clear experimental evidence of the SWNT IMJs and showed that the M-S IMJ has an electronically sharp interface without the localized junction states, whereas a more extended interface and low-energy localized states are found in the M-M IMJ. Based upon their experimental results, they proposed several atomic models to simulate their observed data. Then, the LDOSs were calculated by a simple $\pi$-orbital tight-binding (TB) method, and compared with their experimental observation. It was found that their model I containing three separated 5-7 pairs matched well the experimental data of the (21, -2)/(22, -5) junction. On the other hand, they failed to well describe the M-M (11, 8)/(9, 6) junction, which suggests that more detailed studies, both experimental and theoretical, are still needed to gain further insight on the characteristic features of the M-M IMJ, especially its topological defect structure. Therefore, in this paper, we pay more attention to the different possible geometrical structures of the M-M (11, 8)/(9, 6) IMJ and their effects on the electronic and transport properties of the M-M IMJs. We also calculate changes of the LDOS introduced by a combined 5-7/7-5 defect and show how these changes can be closely related to the defect structure. Compared to those earlier researches, we have included the $\sigma$-$\pi$ hybridization in our TB calculations to see if the $\sigma$ bonds have really important effects on physical properties of IMJs.

This paper is organized as follows. In section 2, several M-M IMJs have been constructed by a projecting method, followed by the structure optimizations. Four-orbital TB computational schemes for the LDOS and transport properties of the M-M IMJs are presented in detail in section 3. Some concluding remarks are offered in section 4.

## II. Formation of the junctions

The geometrical structure of a SWNT can be uniquely determined by a chiral vector on a graphite sheet, $\vec{C} = n\vec{e}_1 + m\vec{e}_2$, where $\vec{e}_1$ and $\vec{e}_2$ are the two primitive lattice vectors of the graphite sheet, and $n$, $m$ are a pair of integers, by which the diameter and helicity of the SWNT can be defined. On the other hand, we can also divide a chiral $(n,m)$ SWNT into a series of connecting closed "zigzag rings" [17]. Each ring contains $n$ steps along $\vec{e}_1$ and $m$ steps along $\vec{e}_2$, and can be visualized as the "unit cell" of the tube, which consists of $k = 2(n+m)$ connected lattice points, labeled 1st, 2nd, …, Kth, respectively. Therefore, each lattice point on the nanotube surface can be described by another pair of integers $(l,k)$, where $l$ is the label of the ring and $k = 1,2,...,K$ represents the lattice point along the *l*th ring. The advantage of such a description for a SWNT is to facilitate the formation of various IMJs.

The first step for constructing an IMJ by addition of pentagon and heptagon defects in the hexagonal lattice is to generate a two-dimensional map of the IMJ structure. We put the divided layers of two different SWNT segments in a plane, making them close to each other as possible as they can. The convolution from this map to the three-dimensional tubular structure can be made by a transformation into the cylinder coordinate in the same axial direction. Of course, the differences in diameter and helicity between the two SWNTs cause the appearance of topological defects on the interface of the IMJ under the constraint of each atom having only threefold coordinates in order to keep the sp$^2$ configuration. Since a zigzag ring in the (11, 8) tube contains eight more atoms than that in the (9, 6) tube, there should be at least two pairs of pentagon and heptagon defects to form the IMJ, where the pentagon rings adjoin the (11, 8) segment and the heptagon rings lie near the (9, 6) half tube. Owing to the Stone-Wales transformation [18], we can change the atomic structure of the junction by rotating a C-C bond around its center to introduce more topological defects and form various connections. The size of an IMJ interface depends on the number and arrangement of the defects in it.

Then, we start a structural relaxation by the combined DFT and the UFF [19-20] calculations to refine the atom coordinates while preserving the connectivity. The optimization processes are carried out on the junction of 2138 atoms, which contains 29 zigzag rings of the (11, 8) tube and 30 zigzag rings of the (9, 6) tube, in addition to a small interface connection between these two tubes. By fixing two zigzag rings at both ends of the junction, we then relax its geometrical structure with the UFF force field, followed by a DFT structural optimization, which is applied to a smaller section containing only 234 atoms cut off from the middle part of the junction. The DFT code used here is Accelrys' DMol3 [21], in which the electronic wave functions are expanded in a double-numeric polarized (DNP) basis set with a real-space cutoff of 4.0 Å. Approximation used in the Hamiltonian is the Harris functional [22] with a local exchange-correlation potential [23].

We have obtained some dozens of the (11, 8)/(9, 6) junctions with different defect structures by this method, among which only six different lowest-energy configurations are illustrated in Fig. 1. It is seen from Fig. 1 that there are smaller structural differences in their junction interfaces, but their electronic properties will differ from each other. Among all the samples we obtained, the model II (a) corresponds to the ground-state structure, which has two 5-7 pairs, one single pentagon and heptagon, all aligned along the axis of the tube. The energy values with respect to the model II (a) for other isomers are given in Table 1, which also summarizes some structural and electronic features for different defect distributions. As indicated in Ref. [15] for the IMJs composed of both zigzag tubes, an IMJ with 5-7 defects parallel to the tubular axis has lower energy than those with the defects along the circumferential direction, which has been clearly corroborated in our calculations for other kind of IMJ structures. Of course, more additive defects will increase the total energy of the junction. For example, a $90^0$ rotation of a C-C bond denoted by an arrow in I (a) can lead to the configuration of III (a) with an additional combined 5-7/7-5 pair defect, which has 3.78meV/atom higher energy than that of I (a). Nevertheless, an appropriate distribution of the defects can decrease more the total energy of the junction, thus overcoming the increase due to the addition of a 5-7 defect. For instance, the atoms in I (a) get about 2meV/atom higher energy than those in II (a) although the latter has one more 5-7 pair defect, which is because all the pair defects tend to parallel the tube axis in the model II (a). As for bond lengths of the junction, our optimization calculation presents a distribution centered at 1.42 Å with a slightly longer bond length in the heptagon while appreciably decreased in the pentagon. It is remarkable that in some IMJ structures, e.g., the models III (b) and III (c), the length of a shared bond between two heptagons in a 7-7 pair defect is much longer than 1.42 Å, reaching about 1.60 Å.

## III. Electronic properties

We have developed an algorithm to study the IMJ's electronic properties by using a tight-binding model involving both *s* and *p* orbitals with only the nearest-neighbor interactions considered. The hopping integrals are taken to be close to those used for graphite [24], which were successfully used to study the SWNT with small radius [25], the SWNT with polygonized cross sections [26], etc. Among the four orbitals per atom, its *s* level is located at $\varepsilon_s^0 = -7.3$ eV below the triply-degenerated *p* level taken as the zero of energy. The hopping parameters for the nearest-neighbor pairs are taken as $V_{ss\sigma} = -4.43$ eV, $V_{sp\sigma} = 4.98$ eV, $V_{pp\sigma} = 6.38$ eV and $V_{pp\pi} = -2.66$ eV. Such hopping integrals are multiplied by a scaling factor $(r_0/r_i)^2$, where $r_0$ and $r_i$ are bond length of the perfect tube and the IMJ, respectively, to include the effects of the distorted bonds in the junction region [27]. A longer junction length of about 100 Å is taken in our calculation in order to avoid quantum size effects. Both sides of the IMJ are connected to perfect (11, 8) and (9, 6) tubes respectively, which are treated as semi-infinitely long. Since early TB calculation on the SWNT IMJ was restricted to electrons, we also perform a TB calculation with only -orbital parameter of $V_{pp\pi} = -2.5$ eV in order to compare directly our results with those in Ref. [16] and show the effect of the σ-π mixing on the junction properties.

The calculated LDOS for a small central part of the (11, 8)/(9, 6) junctions shown in Fig. 1 are illustrated in Fig. 2 for a narrow energy interval around the Fermi level $E_F$ (zero energy). Here, we have averaged the LDOS over each section with length of 3 Å, which is numbered along the common tubule axis and begins from the origin, located in the middle of the interface region. That is to say, the indices (1, 2, 3 and 4) following "L" and "R", which correspond to the (11, 8) and (9,6) side respectively, refer to the specific sections in the "L" and "R" parts away from the interface, just as shown in the middle inset of Fig. 2. Therefore, the LDOS as a function of distance from the interface on either side of the junction can be directly compared to each other. Fig. 2 shows how the LDOS varies from section to section along the (11, 8)/(9, 6) junction, from which common features of the M-M junction can be easily obtained. The LDOS are mostly distorted in the interface region consisting of L1 and R1, and asymmetric around $E_F$ due to existence of the topological defects. The distortions disappear much more swiftly away from the interface into the (9, 6) side than into the (11, 8) segment. This behavior reflects the different screening characteristics of the two segments. Far from the interface region, the DOS features of the perfect tube are progressively recovered. For example, the LDOS curves at L4 and R3 show the basic electronic structures (especially the band gap) of the individual (11, 8) and (9, 6) tubes, respectively.

Now, we pay more attention to the σ-π mixing effect on electronic properties of the IMJ. A TB calculation on the (11, 8)/(9, 6) hybrids has been carried out, not only in the π approximation but also in the σ-π coupling. The π-only results for the model III (b) is very similar to those given by Min Ouyang *et al*. [16], showing a low-energy peak at -0.59 eV. The four-orbital calculation only shifts the position of the localized defect state to -0.64 eV in addition to a very small dip at the Fermi level due to the tube curvature, which is contrary to the experimental observation in Ref. [16]. We may conclude that the σ-π mixing would have a negligible effect on the electronic properties of the SWNT IMJs around Fermi energy, even in the case of two tubes having a large difference in diameters such as the (11, 8) and (9, 6) IMJ. Therefore, in order to interpret properly appearance of TWO localized states below $E_F$ observed experimentally, it is needed to look for more possible configurations for the (11, 8)/(9, 6) junction rather than to include only the σ- π orbital hybridization.

Of special interest in the electronic structures of the junction is emergence of the localized defect states near $E_F$, shown in Fig. 2 as well as presented in Table 1. It is a nontrivial and important problem to know the relationship between these localized defect states and the topological defect structures of the IMJ. It is found by our calculations that for the M-M IMJs, the shared bond between two heptagons in a 7-7 pair defect has larger contribution to the LDOS peaks lying below $E_F$ than the isolated 5-7 pair defects. For example, the junction models I (a) and II (a) have no localized state below $E_F$. However, in the model I (a), although no common edge in a heptagon pair exists, an isolated heptagon lies near the 5-7 pair defect, and can be connected to its heptagon by only one C-C bond, which, consequently, leads to appearance of a smooth convexity around -0.40 eV in the L1 and L2 sections. Furthermore, when we introduce an additional 5-7/7-5 pair into the model I (a) by rotating one bond denoted by an arrow in Fig.2 around its center, the model III (a) is then obtained. Correspondingly, one more shared bond appears between two heptagons in the additive 5-7/7-5 pair defect, denoted by an arrow in the model III (a), and two more shared bonds between more heptagons coming from the rotation. The topological defects of the model III (a) cause appearance of a lower-energy peak at -0.52 eV in its DOS. Such kind of results can also be seen in the M-S junction models connecting (21, -2) and (22, -5) tubes [16]. Therefore, the occupied valence states of the IMJ seem to be mainly affected by presence of the shared bond between two heptagons. However, the number of the defect states below $E_F$ does not equal the number of the shared bonds between the heptagonal defects. By investigating the geometrical and electronic structures in detail, we believe that the angle between the tubular axis and the axis of the 7-7 defects is one of the factors to determine the position of a low-energy peak below $E_F$. The angle is different, so does the position of the peak, and vice versa. For example, in the model III (b), two pairs of 7-7 defects with almost equal inclination respective to the tubular axis cause only one low-energy peak in the valence band of their LDOSs. Which one among the various configurations of the IMJs, shown in Fig. 1, is the best can be decided by comparisons of their LDOSs with the experimental data. It is found in Ref. [16] that the (11, 8)/(9, 6) IMJ has two localized states at -0.55 and -0.27 eV separately. From Fig. 2 and Table 1, it is clearly seen that the junction model III (c) has TWO localized defect states located at –0.68 and – 0.36 eV, respectively, which are well consistent with the experimental values. So, we conclude that the model III (c) probably reflects the real topological structure of the M-M (11, 8)/(9, 6) IMJ used in the experiment [16] more reasonably than any others.

We are also concerned with environment around the defects in the interface of the junction. Our calculated results indicate that the positions of the localized states above $E_F$ may be controlled by both the distribution of the defect pairs and the number of hexagons around the defects. The more hexagons around the defects, the higher the peak energy in the conduction band, which can be also verified by several other configurations of the (11, 8)/(9, 6) matched tubes. As for the number of localized states above $E_F$, it seems that in a lot of cases, e.g., in the model II (b), III (a), III (b), and III (c), the peaks caused by the topological defects appear in pairs in the valence and conduction band. Due to the complicated correlation between the defect structures and the electronic properties, other factors, such as the bond lengths of topological defects, may be also taken into account.

Finally, we simply discuss effects of the topological defects and the σ-π mixing on the conductance of the M-M (11, 8)/(9, 6) IMJ. The coupling of the junction to the left (right) lead, which is simulated by the pure SWNT on left (right) side, can be described by a coupling matrix $h_{LC}$ ($h_{RC}$). The conductance of the model junction is calculated by using the Green's function matching approach [28-29] within the Landauer formalism. As an example, the obtained transmission coefficient $T(E)$ for the model III (c) is given in Fig. 3, incorporating the corresponding results for the perfect (11, 8) and (9, 6) tubes. It is seen from Fig. 3 that $T(E)$ is asymmetric around $E_F$ and smaller than that of either pure (11, 8) or (9, 6) tube due to existence of the potential barrier at the M-M interface. Since the localized states near $E_F$ can scatter electrons, there are several deeper drops in the $T(E)$ at the corresponding energies of the defect states, which offers another experimental method to detect these localized defect states produced by the topological defects in the IMJ interface. On the other hand, it is interesting to note emergence of a sharp peak at -1.36 eV in the four-orbital TB calculation of the model III (c), which is absent in the π-only TB calculation and definitely produced by the σ-π coupling. Since it is located between the first and the second van Hove singularities of the perfect tubes, we may conclude again that the σ-π coupling has a negligible effect on the electronic properties of the IMJs in the energy region around Fermi energy or between the first van Hove singularities of the two perfect tubes forming the IMJs.

## IV. Conclusions

We have studied the structural and electronic properties of the M-M IMJs made of (11, 8) and (9, 6) tubes by introducing the topological defects. Several kinds of the M-M (11, 8)/(9, 6) IMJ are constructed along their common axis first by a new layer-divided technique, and then further structure-optimized by the combined DFT-UFF program. Based upon the experimental results [16], we can determine the most likely geometrical structure among the various IMJs. It is found that the distributions of the topological defects as well as the hexagon number around the defects have important effects on the positions of the localized states. Comparison between our computational results for different junction models suggests that the geometrical distribution of the heptagons in the junction has larger effect on the localized states than that of the isolated 5-7 pairs. It is also indicated that the σ-π coupling effect near Fermi level on the DOS and the electronic properties of the IMJs can be neglected. Finally, the junction model III (c) probably reflects a real atomic structure of the M-M IMJ observed in Ref.[16] because two peaks at -0.36 and -0.68 eV obtained from our calculations agree with those measured in the experiment.

Acknowledgment: This work was supported by the Natural Science Foundation of China under Grant No. 10074026, and No.A040108. The authors acknowledge also support from a Grant for State Key Program of China through Grant No. 1998061407.

Table.1. Structural and electronic features for the six isomers of the (11, 8)/(9, 6) IMJ studied in this paper. Their energies are given with respect to the ground-state configuration (II (a)) found by us. The bond lengths are averaged over the pentagons, and heptagons of the corresponding junctions. The minimum bond length in pentagons and the maximum in heptagons are listed in the parentheses. Peak positions of the localized states around $E_F$ for the different defect distributions of the (11, 8)/(9, 6) IMJ are also given for comparison.

| Model | Energy per atoms (meV/atom) | Bond length in pentagons (Å) | Bond length in heptagons (Å) | Hexagonal Number around the defects | Below $E_F$ (eV) | Above $E_F$ (eV) |
| --- | --- | --- | --- | --- | --- | --- |
| I(a)   | 2.00 | 1.375 (1.344) | 1.459 (1.529) | 18 | -              | 0.63         |
| II(a)  | 0.00 | 1.383 (1.352) | 1.432 (1.479) | 26 | -              | 0.80         |
| II(b)  | 1.93 | 1.378 (1.348) | 1.444 (1.573) | 20 | -0.72          | 0.68         |
| III(a) | 5.78 | 1.388 (1.366) | 1.456 (1.603) | 16 | -0.52          | 0.44         |
| III(b) | 5.19 | 1.380 (1.355) | 1.453 (1.596) | 24 | -0.64          | 0.78         |
| III(c) | 3.22 | 1.392 (1.354) | 1.454 (1.589) | 22 | -0.68 & -0.36  | 0.40 & 0.72  |

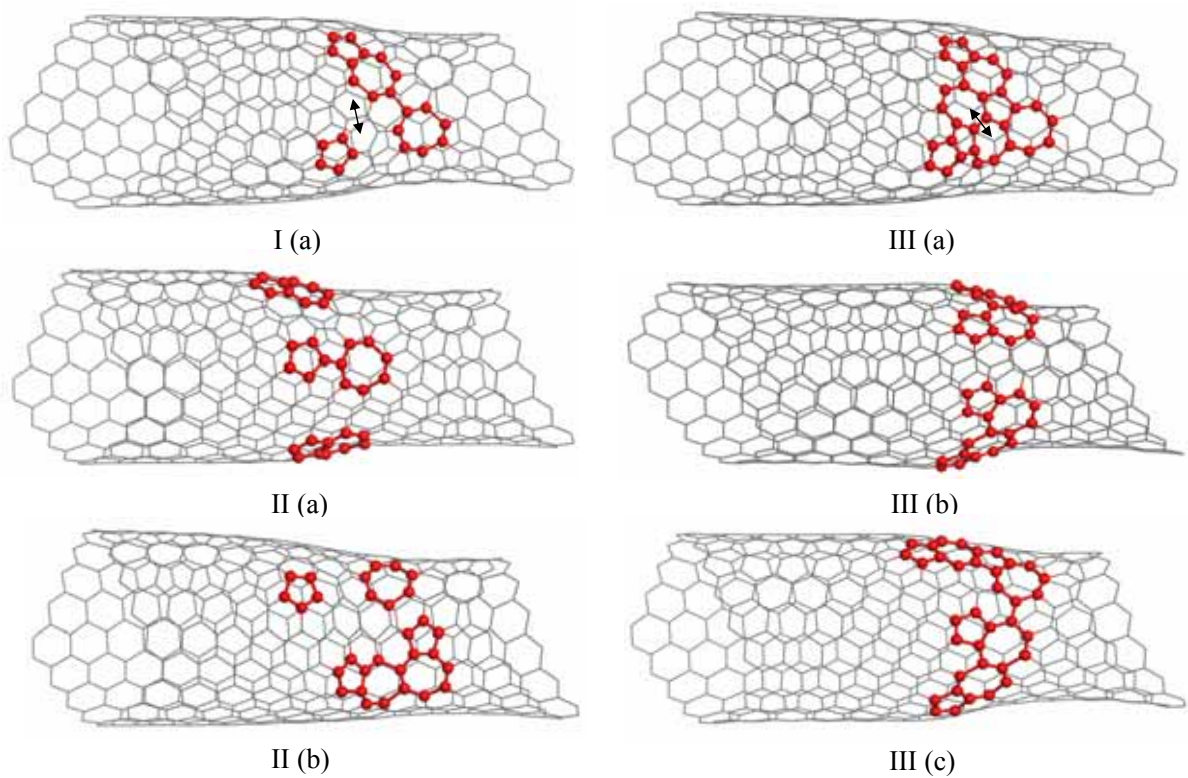

FIG.1. Selected several optimized geometrical structures for the (11, 8)/(9, 6) IMJ with different topological defect configurations, in which the carbon atoms of the pentagons and heptagons are indicated by solid balls. The structures I, II, and III contain a joint section with two, three and four pairs of pentagons and heptagons, respectively. Each of them may have different isomers, such as indicated by (a), (b) or (c).

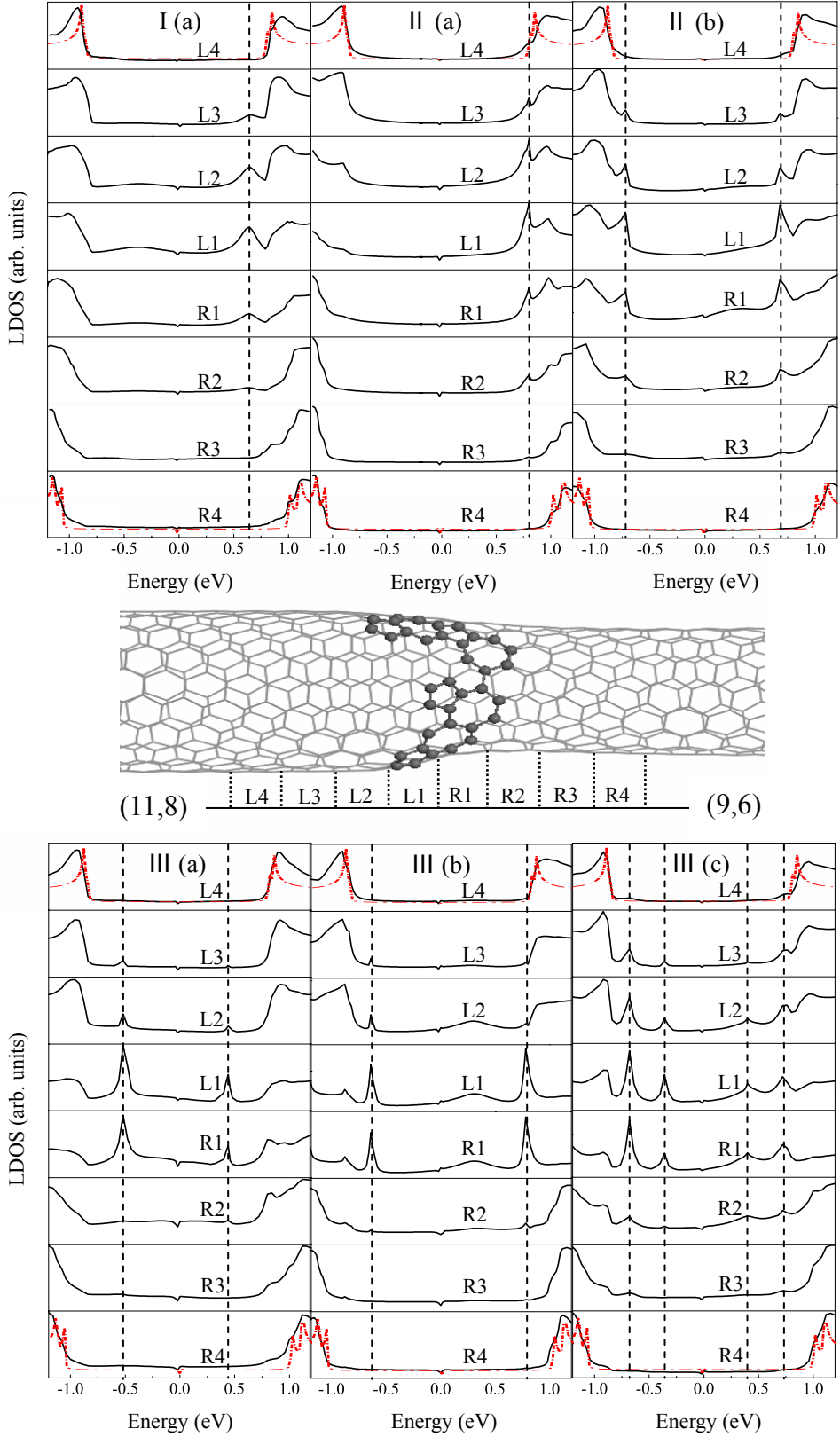

FIG.2. Four-orbital TB calculated LDOSs for the junctions shown in FIG. 1, which are averaged over a section with its length of 3 Å. The horizontal lines indicate the zero-density levels, while the broken vertical lines indicate the positions of the localized defect states in the interface region. The dashed curves in the top and bottom of the figure correspond to the calculated bulk DOSs of the pure (11, 8) and (9, 6) tubes, respectively. The middle panel shows the considered sections schematically.

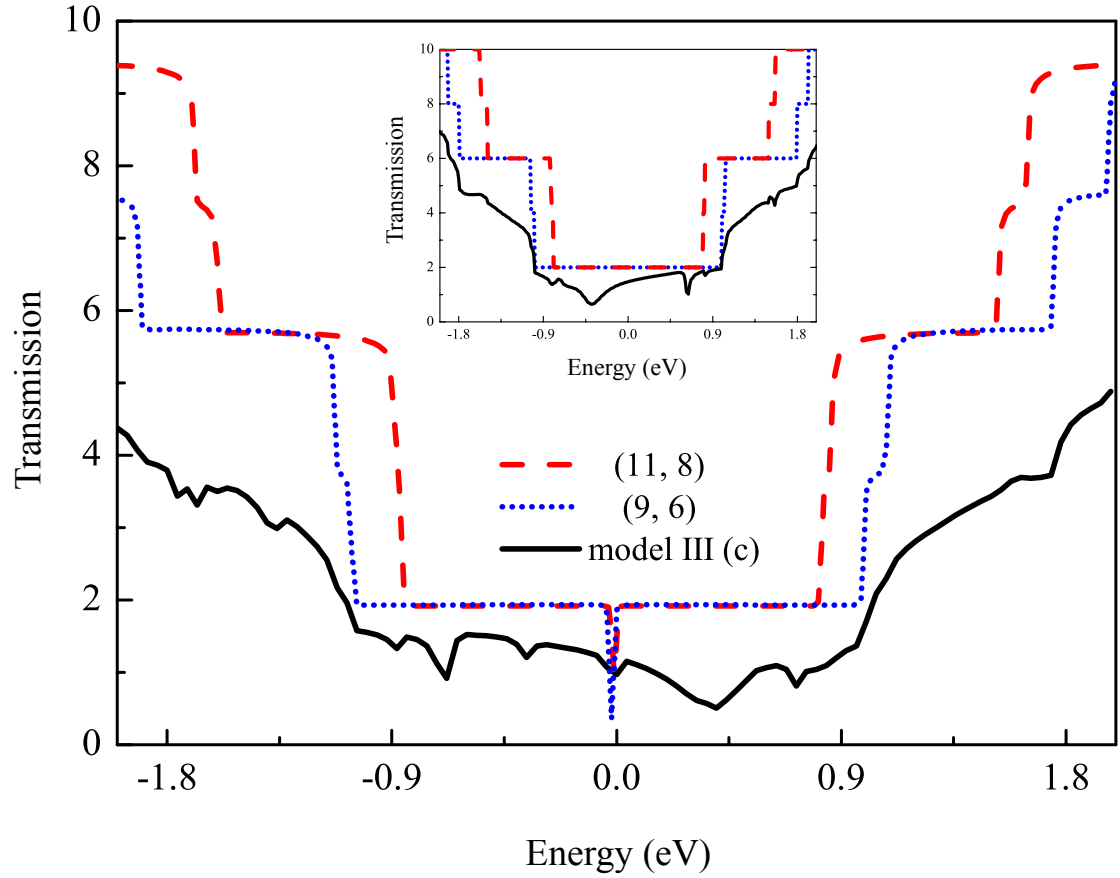

FIG. 3. Transmission coefficient vs energy for the model III (c) and those pure (11, 8) and (9,6) tubes. The inset shows the $\pi$-only TB calculation results.